\documentclass[10 pt, letter]{article}
\usepackage{graphicx}
\usepackage{subfigure}
\usepackage{caption}
\usepackage{amsmath}
\usepackage{amsthm}
\usepackage{enumerate}
\usepackage{subfig}
\usepackage{float}
\usepackage{amsfonts}
\usepackage[spanish]{babel}
\usepackage[latin1]{inputenc}
\usepackage{amssymb}
\usepackage{anysize}
\usepackage{helvet}
\usepackage{setspace}
\usepackage{multirow}
\usepackage{tabularx}
\usepackage{tabulary}
\usepackage{longtable}
\usepackage{float}
\usepackage{mathtools}
\usepackage[none]{hyphenat}

\usepackage{cite}
\title{\bf Charged particle in a flat box with static electromagnetic field and Landau's levels }
\author{ Gustavo V. L\'opez\footnote{gulopez@cencar.udg.mx}, Jorge A. Lizarraga \footnote{jorge.a.lizarraga.b@gmail.com}.
	\\
	\\
 Departamento de F\'{i}sica, Universidad de Guadalajara,\\
 Blvd. Marcelino Garc\'{i}a Barragan y Calzada Ol\'{i}mpica, \\ CP 44200, Guadalajara, Jalisco, M\'exico, \\ 
 \\  
 }
\begin{document}
\maketitle

\begin{abstract}
\noindent
We study the quantization of the motion of a charged particle without spin inside a flat box under a static 
electromagnetic field. Contrary to Landau's solution with constant magnetic field transverse to the box, we found a 
non separable variables solution for the wave function, and this fact remains when static electric field is added. However,
the Landau's Levels appear in all cases.
\end{abstract}
\vskip2pc\noindent
{\bf Key words: } Landau's Levels, quantum Hall effect.
\vskip1pc\noindent
{\bf PACS:} 03.65.-w, 03.65.Ca, 03.65.Ge
\vskip1cm
\newpage
\section{Introduction}
Landau' solution \cite{landau2013quantum} of a charged particle in a flat surface with magnetic field has become of great importance in 
understanding integer hall effect \cite{ando1975theory,klitzing1980new,laughlin1981quantized,halperin1982quantized,laughlin2000fractional}, fractional Hall effect \cite{laughlin2000fractional,tsui1982two,laughlin1983anomalous,jain1989composite}, and topological insulators \cite{bernevig2006quantum,konig2007quantum,li2009topological,jiang2009numerical,groth2009theory,mong2010antiferromagnetic,pesin2010mott}. This last elements
promise to become essential for future nanotechnology devices \cite{fu2008superconducting,freedman2002modular,kitaev2006anyons}.  Due to this considerable application of the
Landau's levels, it is worth to re-study this problem and its variations with an static electric field. In this paper, we show that
there exists a non separable solution for this type of quantum problems, but having the same Landau's levels. In our cases,
instead of having a flat surface, we consider to have a flat box with lengths $L_x$, $L_y$, and $L_z$ such that 
$L_z\ll L_x, L_y$
\section{Analytical approach for the case ${\bf B}=(0,0,B)$}
Let us consider a charged particle ``q" with mass ``m" in a flat box with a constant magnetic field orthogonal to the 
flat surface, ${\bf B}=(0,0,B)$, as shown in the next figure. 
\begin{figure}[H]
\includegraphics[scale=0.12, angle=0]{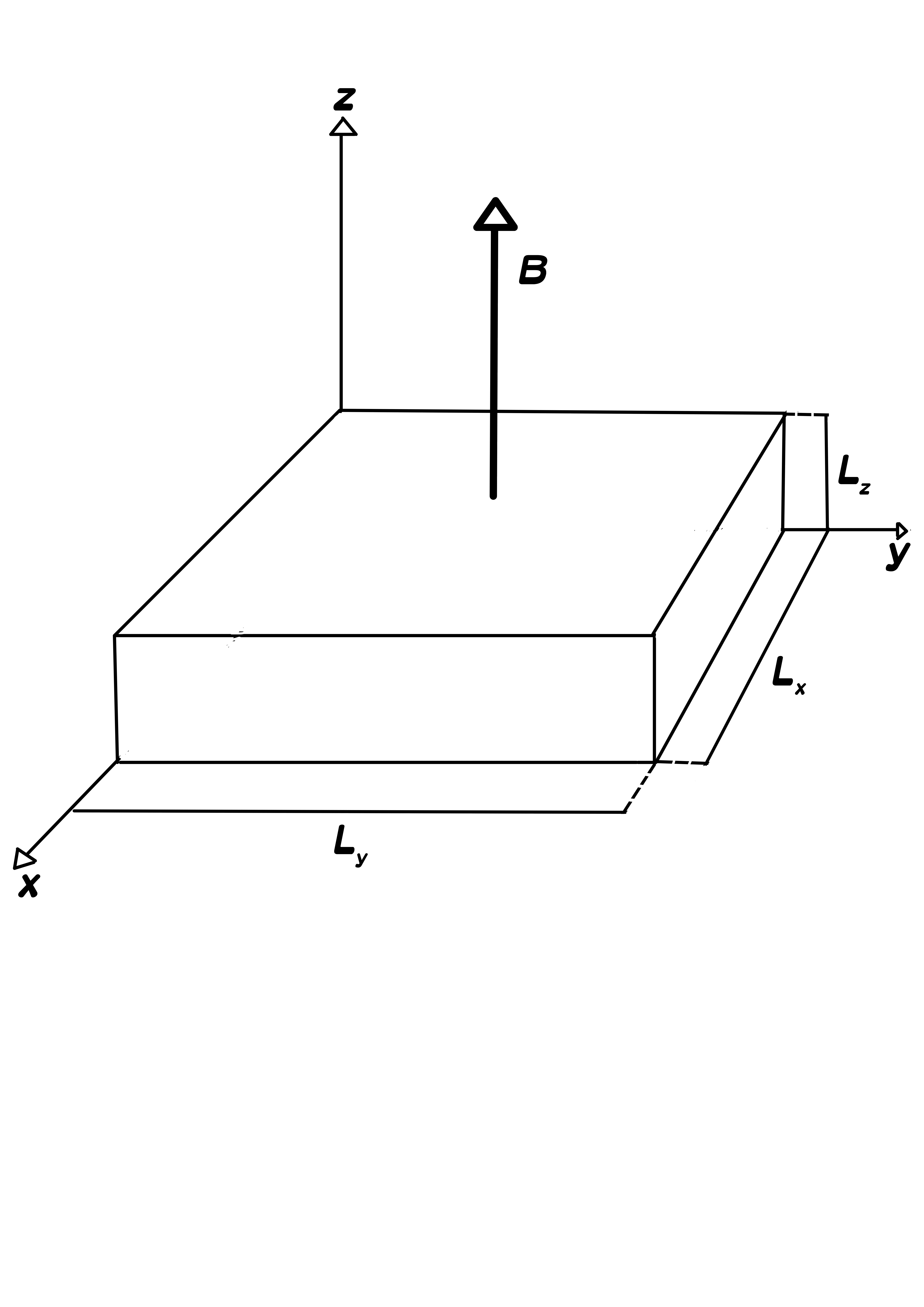}
	\centering
    \caption{Electric charged in a flat box with magnetic field}
\end{figure}
For a non relativistic  charged particle, the Hamiltonian 
of the system (units CGS) is
\begin{equation}
H=\frac{({\bf p}-q{\bf A}/c)^2}{2m},
\end{equation}
where ${\bf p}$ is the generalized linear momentum, ${\bf A}$ is the magnetic potential such that ${\bf B}=\nabla\times {\bf A}$, 
and ``c" is the speed of light. We can choose the Landau's gauge to have the vector potential of the form ${\bf A}=(-By,0,0)$. Therefore,
the Hamiltonian has the following form
\begin{equation}
H=\frac{(p_x+qBy/c)^2}{2m}+\frac{p_y^2}{2m}+\frac{p_z^2}{2m}.
\end{equation}
To quantize the system, we need to solve the Schr\"odinger's equation \cite{messiah1999quantum}
\begin{equation}\label{sch}
i\hbar\frac{\partial\Psi}{\partial t}=\left\{\frac{(\hat{}p_x+qBy/c)^2}{2m}+\frac{\hat{p}_y^2}{2m}+\frac{\hat{p}_z^2}{2m}\right\}\Psi.
\end{equation} 
where $\Psi=\Psi({\bf x},t)$ is the wave function, $\hbar$ is the Plank's constant divided by $2\pi$, $\hat{p}_i$ are the momentum operators
such that $[x_i,\hat{p}_j]=i\hbar\delta_{ij}$. 
Now, the argument used by Landau is that due to commutation relation $[\hat{p}_x, \hat{H}]=0$, between the operators $\hat{p}_x$ and the Hamiltonian $\hat{H}$ 
(implying that $\hat{p}_x$ is a constant of motion), it is possible to replace this component of the momentum by $\hbar k_x$, having a solution for the eigenvalue problem
of  separable variable type, $f_1(t)f_2(x)f_3(y)f_4(z)$. However, we will see that this type of commutation does not imply 
necessarily separability of the solution. Since the Hamiltonian $\hat{H}$ does not depend explicitly on time, the proposition
\begin{equation}
\Psi({\bf x},t)=e^{-iEt/\hbar}\Phi({\bf x})
\end{equation}
reduces the equation to an eigenvalue problem
\begin{equation}\label{eigen1}
\widehat{H}\Phi=E\Phi.
\end{equation}
Then, this equation is written as
\begin{equation}
\left\{\frac{1}{2m}\left(\hat{p}_x^2+\frac{2qB}{c}y\hat{p}_x+\frac{q^2B^2}{c^2}y^2\right)+\frac{\hat{p}_y^2}{2m}+\frac{\hat{p}_z^2}{2m}\right\}\Phi=E\Phi.
\end{equation}
The variable ``z" is separable through the proposition
\begin{equation}
\Phi({\bf x})=\phi(x,y)e^{-ik_zz},\quad\quad k_z\in\Re,
\end{equation}
resulting the following equation
\begin{equation}\label{ns1}
\left\{\frac{1}{2m}\left(\hat{p}_x^2+\frac{2qB}{c}y\hat{p}_x+\frac{q^2B^2}{c^2}y^2\right)+\frac{\hat{p}_y^2}{2m}\right\}\phi=E'\phi,
\end{equation}
where $E'$ is
\begin{equation}
E'=E-\frac{\hbar^2k_x^2}{2m}.
\end{equation}\label{ns2}
That is, the resulting partial differential equation is of the form
\begin{equation}
\frac{1}{2m}\left\{-\hbar^2\frac{\partial^2\phi}{\partial x^2}-i\frac{2qB\hbar}{c}y\frac{\partial\phi}{\partial x}+\frac{q^2B2}{c^2}y^2\phi\right\}-\frac{\hbar^2}{2m}\frac{\partial^2\phi}{\partial y^2}=E'\phi.
\end{equation}
This equation does not admit  a separable variable solution ($\phi(x,y)=f(x)g(y)$) as Landau' solution is, but we can use Fourier transformation \cite{rudin1974fourier} on the variable ``x",
\begin{equation}
\hat{\phi}(k,y)={\cal F}[\phi]=\frac{1}{\sqrt{2\pi}}\int_{\Re}e^{ikx}\phi(x,y)dx,
\end{equation}
to solve this equation. Applying Fourier transformation to this equation, knowing its property  ${\cal F}[\partial\phi/\partial x]=(-ik)\hat{\phi}$, we get
the ordinary differential equation 
\begin{equation}
-\frac{\hbar^2}{2m}\frac{d^2\hat{\phi}}{dy^2}+\frac{m}{2} \omega_c^2(y-y_0)^2\hat{\phi}=E'\hat{\phi},
\end{equation}
where $\omega_c$ is the cyclotron frequency
\begin{subequations}
\begin{equation}\label{cycl}
\omega_c=\frac{qB}{mc}
\end{equation}
and $y_0$ is the displacement parameter
\begin{equation}
y_0=\frac{\hbar c}{qB}k.
\end{equation}
\end{subequations}
This equation is just the quantum harmonic oscillator in the ``y" direction  displaced by a amount $y_0$. So, the solution is
\begin{equation}\label{ho}
\hat{\phi}_n(k,y)=\psi_n(\xi), \quad \xi=\sqrt{\frac{m\omega_c}{\hbar}}(y-y_0), \quad \psi_n(\xi)=A_ne^{-\xi^2}H_n(\xi),
\end{equation}
being $H_n(\xi)$ the Hermit polynomials, and  $A_n$ is a constant of normalization,$An=(m\omega_c/\pi\hbar)^{1/4}/\sqrt{2^n n!}$.
and
\begin{equation}
E'_n=\hbar\omega_c(n+{1}/{2}).
\end{equation}
Now, the solution in the real space  $\phi_n(x,y)$ is gotten by using the inverse Fourier transformation,
\begin{equation}
\phi_n(x,y)={\cal F}^{-1}[\phi_n(k,y)]=\frac{1}{\sqrt{2\pi}}\int_{\Re}e^{-ikx}\psi_n\biggl(\sqrt{\frac{m\omega_c}{\hbar}}(y-\hbar c k/qB)\biggr)dk.
\end{equation} 
Making the change of variable $\sigma=\sqrt{m\omega_c/\hbar}(y-\hbar ck/qB)$, and knowing that the Fourier transformation of the harmonic 
oscillator solution is another harmonic oscillator solution, we get
\begin{equation}
\phi_n(x,y)=\frac{-qB}{\sqrt{mc^2\hbar\omega_c}}e^{-i\frac{qB}{\hbar c}xy}\psi_n\biggl(\frac{qB~x}{\sqrt{mc^2\hbar\omega_c}}\biggr).
\end{equation}
This is indeed the non separable solution of (\ref{ns1}).  Therefore, the normalized eigenfunctions  of the eigenvalue problem (\ref{eigen1}) are (ignoring the sign)
\begin{subequations}
\begin{equation}\label{slv}
\Phi_{n,k_z}({\bf x},t)=\frac{\sqrt{qB}}{\left(mc^2\hbar\omega_c\right)^{1/4}}e^{-i(\frac{qB}{\hbar c}xy-k_zz)}\psi_n\biggl(\frac{qB~x}{\sqrt{mc^2\hbar\omega_c}}\biggr).
\end{equation}
and
\begin{equation}\label{lav}
E_{n,k_z}=\hbar\omega_c(n+\frac{1}{2})+\frac{\hbar^2k_z^2}{2m}.
\end{equation}
\end{subequations}
These eigenvalues represent just the Landau's levels , but its solution (\ref{slv}) is totally different to that given by Landau since it is of non separable type.
Note that there is not displacement at all in the harmonic oscillation solution.  Now, assuming a periodicity in the z-direction,  $\Phi_{n,k_z}({\bf x},t)=\Phi_{n,k_z}(x,y,z+L_z,t)$,
the usual condition $k_zL_z=2\pi n',\quad n'\in {\cal Z}$ makes the eigenvalues to be written as 
and the general solution of Schr\"odinger's equation (\ref{sch}) can be written as
\begin{equation}\label{eig2}
E_{n,n'}=\hbar\omega_c(n+1/2)+\frac{\hbar^22\pi^2}{mL_z^2}n'^2.
\end{equation}
We must observed that this quantum numbers correspond to the degree of freedom in the ``y (n)" and ``z(n')" directions.  
The quantization conditions of the magnetic flux appears rather naturally since by asking periodicity in the y direction $\Psi({\bf x},t)=\Psi(x,y+L_y,z,t)$,
this one must be satisfied for any $x\in[0,L_x]$. So, in particular for $x=L_x$. Thus, it follows from the phase term that
\begin{equation}\label{qflux}
\frac{qB L_xL_y}{\hbar c}=2\pi j,\quad\quad j\in{\cal Z},
\end{equation}
where  $BL_xL_y$ is the magnetic flux crossing the surface with area $L_xL_y$, and  $\hbar c/q$ is the so called quantum flux \cite{bascom1961qfluxs}. Then, equation (\ref{slv}) is 
\begin{equation}\label{slo}
\Phi_{nn'j}({\bf x},t)=\frac{\sqrt{qB}}{\left(mc^2\hbar\omega_c\right)^{1/4}}e^{-i(\frac{2\pi j}{L_xL_y}xy-\frac{2\pi n' }{L_z}z)}\psi_n\biggl(\frac{qB~x}{\sqrt{mc^2\hbar\omega_c}}\biggr).
\end{equation}
The degeneration of the eigenvalues (\ref{eig2}) comes from the degree of freedom in ``x" and 
can be obtained by making use the following quasi-classical argument: given the energy of the harmonic oscillator $E_o=\hbar\omega_c(n+1/2)$,
we know the the maximum displacement of the particle (classically) is given by $x_{max}=\pm\sqrt{2E_o/m\omega_c^2}$, and since the periodicity in the variable `y" mentioned before is valid for any ``x"
value, we must have that the maximum value of the quantum number ``j" must be
\begin{equation}
\Delta j=\frac{qBL_y}{\pi\hbar c}x_{max}=\frac{qBL_y}{\pi\hbar c}\sqrt{\frac{2\hbar(n+1/2)}{m\omega_c}},
\end{equation}  
and this represents the degeneration, $D(n)$, we have in the system
\begin{equation}\label{deg1}
D(n)=\biggl[\frac{qBL_y}{\pi\sqrt{mc^2\hbar\omega_c}}\sqrt{2n+1}\biggr].
\end{equation}
where $[\xi]$ means the integer part of the number $\xi$. 
Therefore, the general solution (absorbing the sign in the constants) is
\begin{equation}
\Psi({\bf x},t)=\sum_{n,'n}\sum_{j=0}^{D(n)}C_{nn'j}\sqrt{\frac{2\pi j}{L_xL_y}}\left(\frac{\hbar}{m\omega_c}\right)^{1/4}
e^{-i(\frac{2\pi j}{L_xL_y}xy-\frac{2\pi n' }{L_z}z)}e^{-i\frac{E_{n,n'}}{\hbar}t}
\psi_n\biggl(\sqrt{\frac{\hbar}{m\omega_c}}\left(\frac{2\pi j}{L_xL_y}\right)x\biggr),
\end{equation}
where the constants $C_{nn'j}$ must satisfy that $\sum_{n,n',j}|C_{nn'j}|^2=1$. The Landau's levels $E_{n,n'}$ are given by expression (\ref{eig2}).
\section{ Analytical approach for the case ${\bf B} \perp{\bf E}$.}
This case is illustrated on the next figure, 
\begin{figure}[H]
\includegraphics[scale=0.12,angle=0]{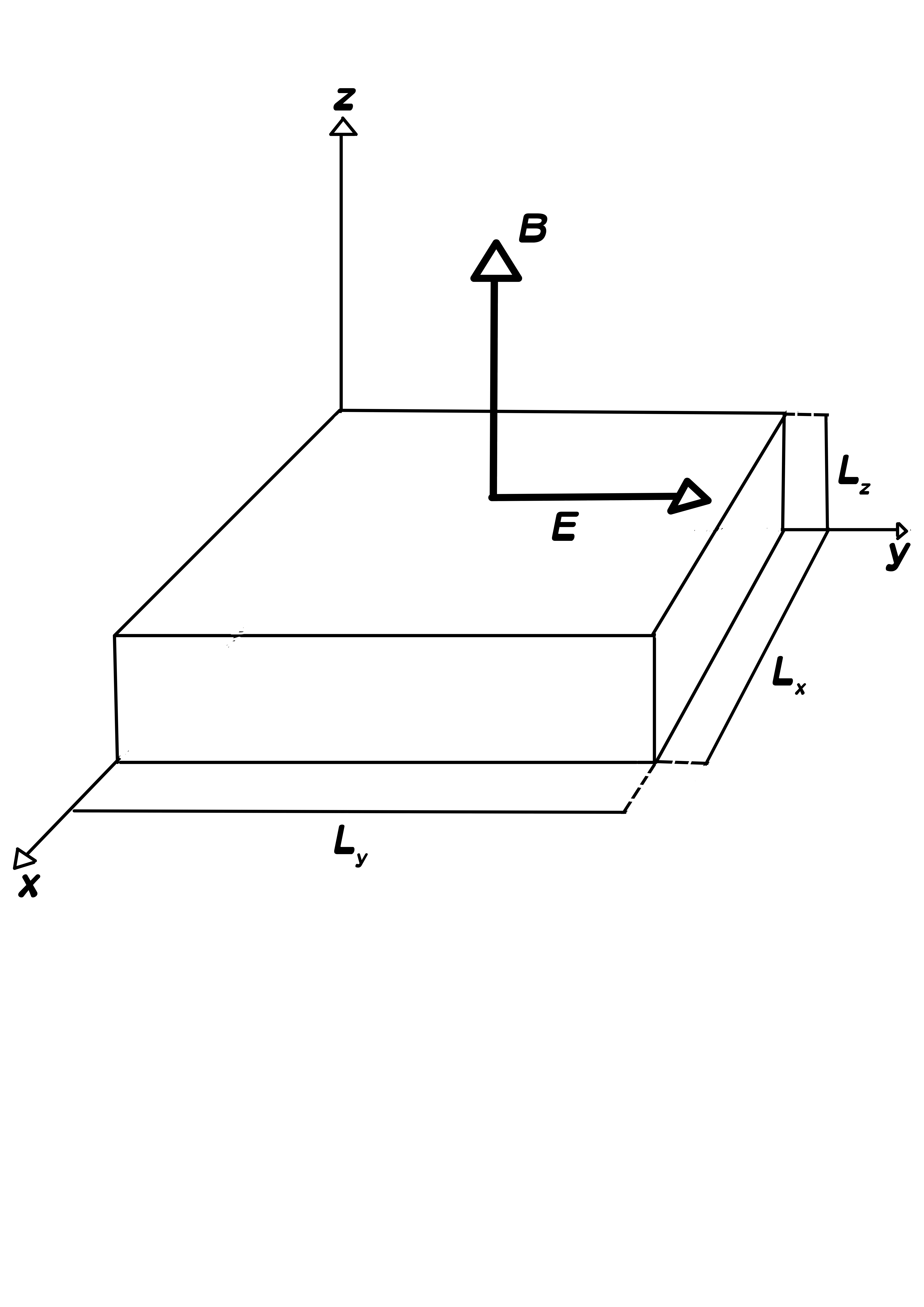}
	\centering
    \caption{Electric charged in a flat box with magnetic and electric fields}
\end{figure}
where the magnetic and electric constant fields are given by ${\bf B}=(0,0,B)$ and ${\bf E}=(0,{\cal E},0)$. 
We select Landau's gauge for the magnetic field such that the vector and scalar potentials are ${\bf A}=(-By,0,0)$ and $\phi=-{\cal E}y$. Then, our Hamiltonian is
\cite{yoshioka2013quantum,datta1997electronic,laughlin1999nobel}
\begin{equation}
\hat{H}=\frac{(\hat{\bf p}-\frac{q}{c}{\bf A})^2}{2m}+q\phi({\bf x},) 
\end{equation} 
and the Schr\"odinger's equation,
\begin{equation}
i\hbar\frac{\partial\Psi}{\partial t}=\hat{H}\Psi,
\end{equation} 
is written as
\begin{equation}\label{sho2}
i\hbar\frac{\partial\Psi}{\partial t}=\left\{\frac{1}{2m}\biggl(\hat{p}_x+\frac{qB}{c}y\biggr)^2+\frac{\hat{p}_y^2}{2m}+\frac{\hat{p}_z^2}{2m}\right\}\Psi-q{\cal E}y\Psi.
\end{equation}
Using the definition $\hat{p}_j=-i\hbar\partial/\partial x_j$ and the commutation relation $[x_k,\hat{p}_j]=i\hbar\delta_jk$, the above expression is written as the following partial differential equation
\begin{equation}
i\hbar\frac{\partial\Psi}{\partial t}=-\frac{\hbar^2}{2m}\frac{\partial^2\Psi}{\partial x^2}-i\frac{qB\hbar}{mc}\frac{\partial\Psi}{\partial x}+\frac{q^2B^2}{2mc^2}y^2\Psi
-\frac{\hbar^2}{2m}\frac{\partial^2\Psi}{\partial y^2}-\frac{\hbar^2}{2m}\frac{\partial^2\Psi}{\partial z^2}-q{\cal E}y\Psi.
\end{equation}
Taking the Fourier transformation with respect the x-variable, $\hat{\Psi}(k,y,z,t)={\cal F}_x[\Psi({\bf x},t)]$, the resulting expression is
\begin{equation}
i\hbar\frac{\partial\hat\Psi}{\partial t}=\left[\frac{\hbar^2k^2}{2m}-\biggl(\frac{qB\hbar k}{mc}+q{\cal E}\biggr)y+\frac{q^2B^2}{2mc^2}y^2\right]\hat{\Psi}-\frac{\hbar^2}{2m}\frac{\partial^2{\hat{\Psi}}}{\partial y^2}
-\frac{\hbar^2}{2m}\frac{\partial^{2}\hat{\Psi}}{\partial z^{2}}.
\end{equation}
By proposing a solution of the form
\begin{equation}
\hat{\Psi}(k,yz,t)=e^{-iE t/\hbar+ik_zz}\Phi(k,y)
\end{equation}
and after some rearrangements, the resulting equation for $\Phi$ is
\begin{equation}
-\frac{\hbar^2}{2m}\frac{d^2\Phi}{dy^2}+\frac{1}{2}m\omega_c^2(y-y_0)^2\Phi=E'\Phi,
\end{equation}
where $\omega_c$ is the cyclotron frequency (\ref{cycl}), and  we have made the definitions
\begin{equation}
y_0=\frac{\hbar c}{qB}k+\frac{mc^2{\cal E}}{qB^2}
\end{equation}
and
\begin{equation}
E'=E-\frac{\hbar^2k^2}{2m}-\frac{\hbar^2k_z^2}{2m}+\frac{1}{2m}(\hbar k+\frac{mc{\cal E}}{B})^2.
\end{equation}
This equation is again the quantum harmonic oscillator on the variable ``y" with a cyclotron frequency $\omega_c$  and displaced by a quantity $y_0$. Therefore, the solution (\ref{ho}) is  
\begin{equation}
\Phi(k,y)=\psi_n\biggl(\sqrt{\frac{m\omega_c}{\hbar}}(y-y_0)\biggr)
\end{equation} 
and
\begin{equation}
E'_n=\hbar\omega_c(n+1/2).
\end{equation}
Thus, the solution in the Fourier space is
\begin{equation}
\hat{\Psi}(k,y,z,t)=e^{-iE_{n,k_z}t/\hbar+ik_zz}\psi_n\biggl(\sqrt{\frac{m\omega_c}{\hbar}}(y-y_0)\biggr)
\end{equation}
with the energies $E_{n,k_z}$ given by
\begin{equation}
E_{n,k_z}=\hbar\omega_c(n+1/2)+\frac{\hbar^2k_z^2}{2m}-\frac{mc^2{\cal E}^2}{2B^2}-\frac{c{\cal E}\hbar}{B}k.
\end{equation}
The solution in the space-time is obtained by applying the inverse Fourier transformation,
\begin{equation}
\Psi_{n,k_z}({\bf x},t)={\cal F}[\hat{\Psi}_{n,k_z}(k,y,z,t)]=\frac{1}{\sqrt{2\pi}}\int_{\Re}e^{-ixk}\hat{\Psi}_{n,k_z}(k,y,z,t)dk,
\end{equation}
which after a proper change of variable and rearrangements , we get the normalized function (ignoring the sign)
\begin{equation}\label{eq39}
\Psi_{n,k_z}({\bf x},t)=\frac{\sqrt{qB}}{\left(mc^2\hbar\omega_c\right)^{1/4}}e^{-i\phi_{n,k_z}({\bf x},t)}\psi_n\biggl(\frac{qB}{\sqrt{mc^2\hbar\omega_c}}\bigl(x-\frac{c{\cal E}t}{B}\bigr)\biggr),
\end{equation}
where the phase $\phi_{n,k_z}({\bf x},t)$ has been defined as
\begin{equation}
\phi_{n,k_z}({\bf x},t)=\left[\hbar\omega_c(n+1/2)+\frac{\hbar^2k_z^2}{2m}-\frac{mc^2{\cal E}^2}{2 B^2}\right]\frac{t}{\hbar}-k_{z}z+\frac{qB}{\hbar c}\left(x-\frac{c{\cal E}t}{B}\right)\left(y-\frac{mc^2{\cal E}}{qB^2}\right).
\end{equation}
asking for the periodicity with respect the variable ``z", $\Psi_{n,k_z}({\bf x},t)=\Psi_{n,k_z}(z,y,z+L_z,t)$, it follows that $k_zL_z=2\pi n'$ where $n'$ is an integer number , and the above phase is 
now written as
\begin{equation}
\phi_{nn'}({\bf x},t)=\left[\hbar\omega_c(n+1/2)+\frac{\hbar^{2}2\pi^{2} n'^{2}}{mL_z^2}-\frac{mc^2{\cal E}^2}{2 B^2}\right]\frac{t}{\hbar}-\frac{2\pi n'}{L_z}z+\frac{qB}{\hbar c}\left(x-\frac{c{\cal E}t}{B}\right)\left(y-\frac{mc^2{\cal E}}{qB^2}\right).
\end{equation}
Note from this expression that the term $e^{-i\phi({\bf x},t)}$ contains the element $e^{i\frac{qB}{\hbar c}xy}$, and by assuming the periodic condition $\Psi({\bf x},t)=\Psi(x,y+L_{y},z,t)$, 
will imply that $\Psi({\bf x},t)$ will be periodic with respect the variable ``y", for any ``x" at any time ``t." In particular, this will be true for $x=L_x$. This bring about the quantization of the magnetic flux of the form
\begin{equation}
\frac{qBL_xL_y}{\hbar c}=2\pi j, \quad J\in{\cal Z}\ ,
\end{equation} 
obtaining the same expression as (\ref{qflux}), and this phase is now depending of the quantum number ``j"
\begin{equation}
\phi_{nn'j}({\bf x},t)=e_{nn'}t/\hbar-\frac{2\pi n'}{L_z}z+\frac{2\pi j}{L_xL_y}xy
-\frac{2\pi j}{L_xL_y}\left[\frac{mc^2{\cal E}}{qB^2}x+\frac{c{\cal E}}{B}ty\right].
\end{equation}
where $e_{nn'}$ is the energy associated to the system,
\begin{equation}
e_{n,n'}=\hbar\omega_c(n+1/2)+\frac{2\pi^2\hbar^2}{mL_z^2}n'^2+\frac{mc^2{\cal E}^2}{2 B^2}.
\end{equation}
In this way, from these relations and the expression (\ref{eq39}) we have a family of solutions $\{\Psi_{nn'j}({\bf x},t)\}_{n,n',j\in{\cal Z}}$ of the Schr\"odinger equation (\ref{sho2}), 
\begin{equation}
\Psi_{nn'j}({\bf x},t)=\sqrt{\frac{2\pi j}{L_xL_y}}\left(\frac{\hbar}{m\omega_c}\right)^{1/4} e^{-i\phi_{nn'j}({\bf x},t)}
\psi_n\biggl(\sqrt{\frac{\hbar}{m\omega_c}}\left(\frac{2\pi j}{L_xL_y}\right)\bigl(x-\frac{c{\cal E}t}{B}\bigr)\biggr),
\end{equation}
Now, by the same arguments we did in the previous case, the degeneration of the systems would be given by (\ref{deg1}), and the general solution would be of the form
\begin{equation}
\Psi({\bf x},t)=\sum_{n,n'}\sum_{j=0}^{D(n)}\widetilde{C}_{nn'j}\Psi_{nn'j}({\bf x},t).
\end{equation}
\newpage
\section{ Analytical approach for the case ${\bf B}\parallel {\bf E}$.}
The following figure shows this case. 
\begin{figure}[H]
\includegraphics[scale=0.12,angle=0]{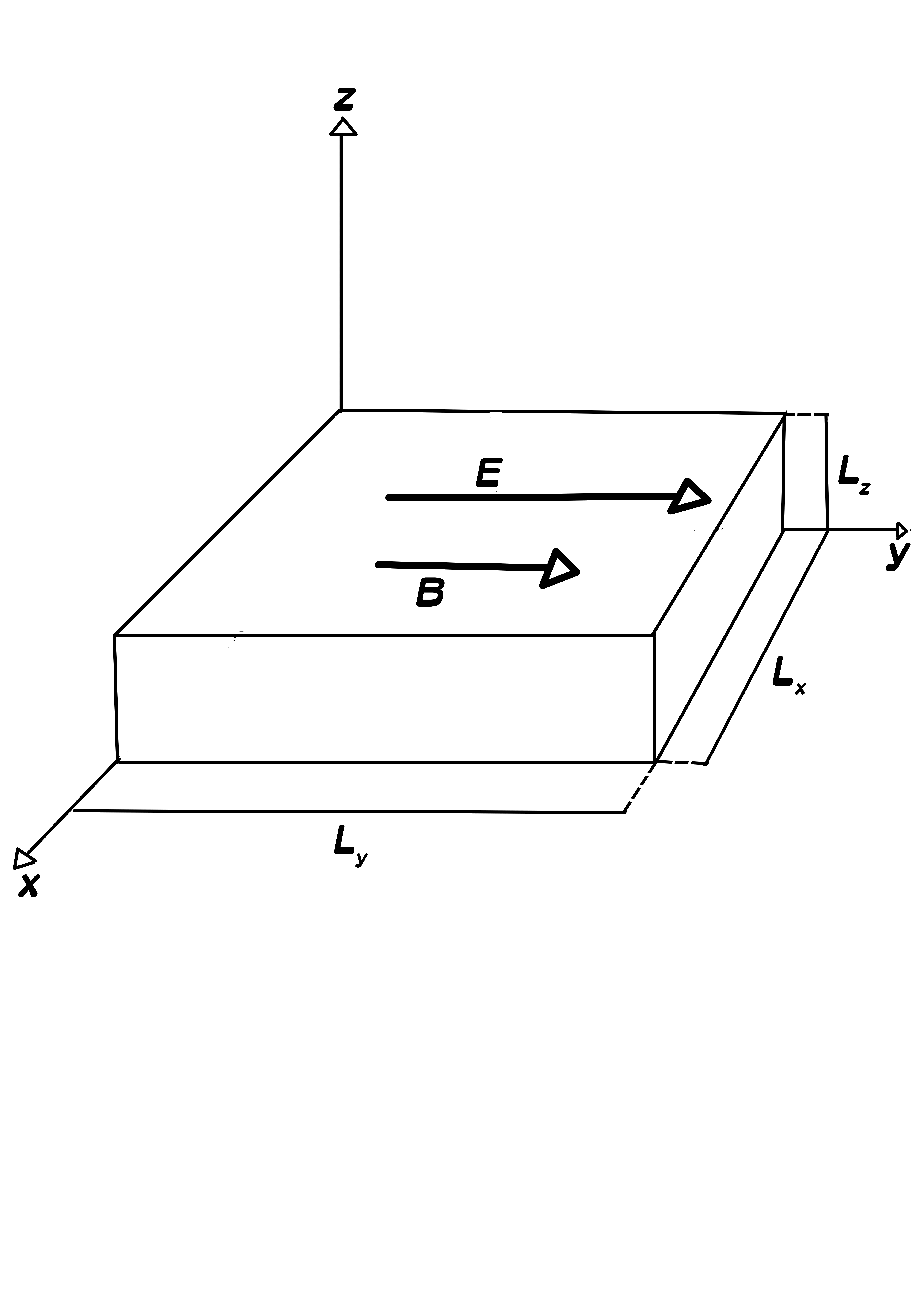}
	\centering
    \caption{Electric charged in a flat box with parallel electric and magnetic fields}
\end{figure}
The fields are of the form ${\bf B}=(0,B,0)$ and ${\bf E}=(0,{\cal E},0)$. The scalar and vector potentials are chosen as
${\bf A}=(Bz,0,0)$ and $\phi=-{\cal E}y$. The Shr\"odinger equation is for this case as
\begin{equation}
i\hbar\frac{\partial\Psi}{\partial t}=\left\{\frac{(\hat{p}_x-qB z/c)^2}{2m}+\frac{\hat{p}_y^2}{2m}+\frac{\hat{p}_z^2}{2m}-q{\cal E}y\right\}\Psi, 
\end{equation}
which defines the following partial differential equation
\begin{equation}\label{sho3}
i\hbar\frac{\partial\Psi}{\partial t}=-\frac{\hbar^2}{2m}\frac{\partial^2\Psi}{\partial x^2}
+i\frac{qB\hbar z}{mc}\frac{\partial\Psi}{\partial x}+\frac{q^2B^2}{2mc^2}z^2\Psi-\frac{\hbar^2}{2m}\frac{\partial^2\Psi}{\partial y^2}-\frac{\hbar^2}{2m}\frac{\partial^2\Psi}{\partial z^2}-q{\cal E}y\Psi.
\end{equation}
Proposing a solution of the form $\Psi({\bf x},t)=e^{-iEt/\hbar}\Phi({\bf x})$, we get the following eigenvalue problem
\begin{equation}
E\Phi=-\frac{\hbar^2}{2m}\frac{\partial^2\Phi}{\partial x^2}
+i\frac{qB\hbar z}{mc}\frac{\partial\Phi}{\partial x}+\frac{q^2B^2}{2mc^2}z^2\Phi-\frac{\hbar^2}{2m}\frac{\partial^2\Phi}{\partial y^2}-\frac{\hbar^2}{2m}\frac{\partial^2\Phi}{\partial z^2}-q{\cal E}y\Phi.
\end{equation}
Applying the Fourier transformation over the x-variable, $\hat{\Phi}(k,y,z)={\cal F}_x[\Phi({\bf x})]$, the following equation arises after some rearrangements
\begin{equation}
E\hat{\Phi}=\frac{(\hbar k+qB z/c)^2}{2m}\hat{\Phi}-\frac{\hbar^2}{2m}\frac{\partial^2\hat{\phi}}{\partial z^2}-\frac{\hbar^2}{2m}\frac{\partial^2\hat{\Phi}}{\partial y^2}-q{\cal E}y\hat{\Phi},
\end{equation}
which can be written as
\begin{subequations}
\begin{equation}
-\frac{\hbar^2}{2m}\frac{\partial^2\hat{\Phi}}{\partial z^2}+\frac{1}{2}m\omega_c(z+z_0)^2\hat{\Phi}-\frac{\hbar^2}{2m}\frac{\partial^2\hat{\Phi}}{\partial y^2}-q{\cal E}y\hat{\Phi},
\end{equation}
where $\omega_c$ is the cyclotron frequency (\ref{cycl}), and $z_0$ has been defined as
\begin{equation}
z_0=\frac{\hbar c}{qB}k.
\end{equation}
\end{subequations}
This equation admits a variable separable approach since by the proposition  $\hat{\Phi}(k,y,z)=f(k,z)g(y)$, the following equations are bringing about
\begin{subequations}
\begin{equation}
-\frac{\hbar^2}{2m}\frac{d^2f}{dz^2}+\frac{1}{2}m\omega_c^2(z+z_0)^2=E^{(1)}f
\end{equation}
and
\begin{equation}
-\frac{\hbar^2}{2m}\frac{d^2g}{dy^2}-g{\cal E}yg=E^{(2)}g,
\end{equation}
\end{subequations}
where $E=E^{(1)}+E^{(2)}$. The solutions of these equations are, of course, the quantum harmonic oscillator and the quantum bouncer, which are given by
\begin{subequations}
\begin{equation}
f_n(k,z)=A_ne^{-\xi^2/2}H_n(\xi), \quad\quad \xi=\sqrt{\frac{m\omega_c}{\hbar}}(z+z_0), \quad\quad E_n^{(1)}=\hbar\omega_c(n+1/2).
\end{equation}
and
\begin{equation}
g_{n'}(y)=\frac{Ai(\tilde\xi-\tilde\xi_{n'})}{|Ai'(-\tilde\xi_{n'})|},\quad\quad \tilde\xi=y/l, \quad \quad E_n^{(2)}=-q{\cal E}l\tilde\xi_{n'},
\end{equation}
\end{subequations}
where $A_n=(m\omega_c/\pi\hbar)^{1/4}/\sqrt{2^nn!}$, $l=(\hbar^2/(-2mq{\cal E}))^{1/3}$,  $Ai(-\tilde\xi_{n'})=0$, and $Ai'(\xi)$ is the differentiation of the Airy function.
In this way, we have
\begin{equation}\label{Fen}
\hat{\Phi}_{n,n'}(k,y,z)=a_{n'}\psi_n\biggl(\sqrt{\frac{m\omega_c}{\hbar}}(z+z_0)\biggr)Ai(l^{-1}(y-y_{n'})),\quad\quad E_{n,n'}=\hbar\omega_c(n+1/2)-q{\cal E}y_{n'},
\end{equation}
where we have defined $a_{n'}$ as $a_{n'}=1/|Ai'(-l^{-1}y_{n'})|$. Now, the inverse Fourier transformation will affect only the quantum harmonic oscillator function $\psi_n$ 
through the k-dependence on the parameter $z_0$, and the resulting expression is
\begin{equation}\label{solpall}
\Phi_{n,n'}({\bf x})=\frac{a_{n'} qB}{\sqrt{mc^2\hbar\omega_c}}e^{i\frac{qB}{\hbar c} xz}\psi_n\biggl(\frac{qB x}{\sqrt{mc^2\hbar\omega_c}}\biggr)Ai\bigl(l^{-1}(y-y_{n'})\bigr).
\end{equation} 
Now, asking for the periodicity condition of the above solution with respect the z-variable, $\Psi({\bf x},t)=\Psi(x,y,z+L_z,t)$, the periodicity must satisfy  for any x-values, and in particular for $x=L_x$. Thus
it follows the quantization expression for the magnetic flux 
\begin{equation}
\frac{qBL_xL_z}{\hbar c}=2\pi j, \quad\quad j\in{\cal Z}.
\end{equation} 
Using the same arguments shown above for the degeneration of the system, we have the same expression (\ref{deg1}) for the degeneration of the system 
and the function (\ref{solpall}) is given by (normalized) 
\begin{equation}\label{soo}
\Phi_{nn'j}({\bf x})=a_{n'}\sqrt{\frac{2\pi j}{L_xL_y}}\left(\frac{\hbar}{m\omega_c}\right)^{1/4}e^{i\frac{2\pi j}{L_xL_z} xz}
\psi_n\biggl(\sqrt{\frac{\hbar}{m\omega_c}}\left(\frac{2\pi j}{L_xL_y}\right)x\biggr)Ai\bigl(l^{-1}(y-y_{n'})\bigr).
\end{equation} 
Then, we have obtained a family of solution of the Schr\"odinger equation (\ref{sho3}),
\begin{equation}
\Psi_{n,n'}({\bf x},t)=e^{-i E_{n,n'} t/\hbar}\Phi_{nn'j}({\bf x}),
\end{equation}
where the energies $E_{n,n'}$ are given by the expression (\ref{Fen}). The general solution of (\ref{sho3}) can be written as
\begin{equation}
\Psi({\bf x},t)=\sum_{n,n'}\sum_{j=0}^{D(n)}C_{n,n'}^* e^{-i E_{n,n'} t/\hbar} e^{i\frac{2\pi j}{L_xL_z} xz}\tilde{u}_{n,n'}(x,y),
\end{equation}
with the condition $\sum_{n,n'}|C_{n,n'}^*|^2=1$, and where it has been defined the functions $\tilde{u}_{n,n'}$ as
\begin{equation}
\tilde{u}_{n,n'}(x,y)=a_{n'}\sqrt{\frac{2\pi j}{L_xL_y}}\left(\frac{\hbar}{m\omega_c}\right)^{1/4}
\psi_n\biggl(\sqrt{\frac{\hbar}{m\omega_c}}\left(\frac{2\pi j}{L_xL_y}\right)x\biggr)Ai\bigl(l^{-1}(y-y_{n'})\bigr).
\end{equation}

\subsection{Same system but with new magnetic gauge.}
Let us consider the magnetic gauge given such that the vector potential is of the form ${\bf A}=(0,0,-Bx)$,
and the potential is the same $\phi=-{\cal E}y$. Passing directly to the eigenvalue problem for the Schr\"odinger equation when we select the wave function of the form $\Psi({\bf x},t)=e^{-iEt/\hbar}\Phi({\bf x})$, the resulting
equation is
\begin{equation}
-\frac{\hbar^2}{2m}\frac{\partial^2\Phi}{\partial x^2}-\frac{\hbar^2}{2m}\frac{\partial^2\Phi}{\partial y^2}-\frac{\hbar^2}{2m}\frac{\partial^2\Phi}{\partial z^2}-i\frac{qB\hbar}{mc}x\frac{\partial\Phi}{\partial z}
+\frac{q^2B^2}{2mc^2}x^2\Phi-q{\cal E}y\Phi=E\Phi.
\end{equation}
Taking the Fourier transformation with respect the z-variable, $\hat{\Phi}(x,y,k)={\cal F}_z[\Phi({\bf x})]$, and making some rearrangements, it follows that
\begin{equation}
-\frac{\hbar^2}{2m}\frac{\partial^2\hat{\Phi}}{\partial x^2}+\frac{1}{2m}\bigl(\hbar k-\frac{qB}{c}x\bigr)^2\hat{\Phi}-\frac{\hbar^2}{2m}\frac{\partial^2\hat{\Phi}}{\partial y^2}-q{\cal E}y\hat{\Phi}=E\hat{\Phi}.
\end{equation} 
This equation admits a variable separable solution of the form $\hat{\Phi}(x,y,k)=\phi_1(k,x)\phi_2(y)$, where the functions $\phi_1$ and $\phi_2$ satisfy the equations
\begin{equation}
-\frac{\hbar^2}{2m}\frac{d^2\phi_1}{dx^2}+\frac{(\hbar k-\frac{qB}{c} x)^2}{2m}\phi_1=E^{(1)}\phi_1
\end{equation}
and
\begin{equation}
-\frac{\hbar^2}{2m}\frac{\partial^2 \phi_2}{\partial y^2}-q{\cal E} y\phi_2=E^{(2)}\phi_2,
\end{equation}
where $E=E^{(1)}+E^{(2)}$. The solution of these equations are
\begin{equation}
\phi_{1n}(k,x)=\psi_n(\xi)=A_ne^{-\xi^2/2}H_n(\xi), \quad \xi=\sqrt{\frac{m\omega_c}{\hbar}}(x-x_0),\quad\quad E_n^{(1)}=\hbar\omega_c(n+1/2)
\end{equation}
and
\begin{equation}
\phi_{2n'}(y)=a_{n'}Ai(l^{-1}(y-y_{n'})),\quad l=\left(\frac{\hbar^2}{-2mq{\cal E}}\right)^{1/3},\quad\quad E_{n'}^{(2)}=-q{\cal E}y_{n'},
\end{equation}
where $\omega_c$ is the cyclotron frequency (\ref{cycl}), $x_0$ is the displacement $x_0=\hbar c k/qB$, $a_{n'}=1/|Ai'(l^{-1}y_{n'})|$ is a constant, and $A_n$ the constant associated to the quantum harmonic oscillator solution. The inverse Fourier transformation affect only the function $\phi_1$, and we have
\begin{equation}
\phi_{1n}(z,x)={\cal F}^{-1}[\phi_{1n}(k,x)]=\frac{-qB}{\sqrt{mc^2\hbar\omega_c}}e^{-i\frac{qB}{\hbar c}xz}\psi_n\biggl(\frac{qB z}{\sqrt{mc^2\hbar\omega_c}}\biggr).
\end{equation}
The periodic condition on the variable ``x", $\Psi({\bf x},t)=\Psi(x+L_x,y,z,t)$, for any value of the other variables, implies that this will happen in particular for the value of $z=L_z$. So, we get
the quantization of the magnetic flux ($BL_xL_y$),
\begin{equation}
\frac{qBL_xL_z}{\hbar c}=2\pi j, \quad\quad j\in {\cal Z}.
\end{equation}
Thus, we have a family of solutions $\{\Psi_{nn'j}({\bf x},t)\}$ of the Shcr\"odinger equation of the form
\begin{equation}
\Psi_{nn'j}({\bf x},t)=e^{-iE_{n,n'}t/\hbar}\Phi_{nn'j}({\bf x}),
\end{equation}
or (normalized and ignoring the sign)
\begin{equation}
\Psi_{nn'j}({\bf x},t)=a_{n'}\sqrt{\frac{2\pi j}{L_xL_y}}\left(\frac{\hbar}{m\omega_c}\right)^{1/4}e^{-i(E_{n,n'}\frac{t}{\hbar}+\frac{2\pi j}{L_xL_z}xz)}
\psi_n\biggl(\sqrt{\frac{\hbar}{m\omega_c}}\left(\frac{2\pi j}{L_xL_y}\right) z\biggr)Ai(l^{-1}(y-y_{n'})).
\end{equation}
\noindent 
By the same arguments about the degenerationn of the systems,  the general solution is just a combination of all of these,
\begin{equation}
\Psi({\bf x},t)=\sum_{n,n'}A_{nn'j}e^{-i(E_{n,n'}\frac{t}{\hbar}+\frac{2\pi j}{L_xL_z}xz)} v_{nn'j}(y,z),
\end{equation}
where the condition $\sum_{n,n'}|A_{nn'j}|^2=1$  must be satisfied, and the function $v_{nn'j}$ is given by
\begin{equation}
v_{nn'j}(y,z)= a_{n'}\sqrt{\frac{2\pi j}{L_xL_y}}\left(\frac{\hbar}{m\omega_c}\right)^{1/4}
\psi_n\left(\sqrt{\frac{\hbar}{m\omega_c}}\left(\frac{2\pi j}{L_xL_y}\right) z\right) Ai\big(l^{-1}(y-y_{n'})\bigr).
\end{equation}  

\section{ Conclusions and comments}
We have studied the quantization of a charged particle in a flat box and under constants magnetic and electric fields for several cases and have shown 
that a full separation of variable solution is not admitted in these cases (contrary to Landau's solution in one of these cases). This situation arises since the commutation 
of  a component of the generalized  linear momentum operator with the Hamiltonian of the system does not imply necessarily that a variable separation of its 
associated variable must exist in the Schr\"odinger equation. However, using the Fourier transformation, we were be able to find the full solution of the problems. 
As expected, Landau's level appears in all these cases, and a characteristic phase which help us to find the quantization of the magnetic flux in a natural way. We consider that
the approach given here maybe very useful to understand quantum Hall effect and related phenomena.    

\bibliographystyle{unsrt} 
 \bibliography{bibliografia}

\end{document}